\documentclass[sigconf]{acmart}
\AtBeginDocument{%
  }

\copyrightyear{2025}
\acmYear{2025}
\setcopyright{acmlicensed}\acmConference[CIKM '25]{Proceedings of the 34th ACM International Conference on Information and Knowledge Management}{November 10--14, 2025}{Seoul, Republic of Korea}
\acmBooktitle{Proceedings of the 34th ACM International Conference on Information and Knowledge Management (CIKM '25), November 10--14, 2025, Seoul, Republic of Korea}
\acmDOI{10.1145/3746252.3761244}
\acmISBN{979-8-4007-2040-6/2025/11}

\usepackage{multirow} 
\usepackage{array}
\usepackage[noend]{algpseudocode}
\usepackage{algorithmicx,algorithm}
\usepackage{tcolorbox}
\usepackage{float}
\usepackage{CJKutf8}

\begin{document}
\begin{CJK}{UTF8}{gbsn}

\title{A Node-Aware Dynamic Quantization Approach for Graph Collaborative Filtering}



\author{Lin Li}
\affiliation{%
  \institution{Wuhan University of Technology}
  \city{Wuhan}
  \country{China}}
\email{cathylilin@whut.edu.cn}

\author{Chunyang Li}
\affiliation{%
  \institution{Wuhan University of Technology}
  \city{Wuhan}
  \country{China}
}
\email{l-cy@whut.edu.cn}

\author{Yu Yin}
\affiliation{%
  \institution{Wuhan University of Technology}
  \city{Wuhan}
  \country{China}
}
\affiliation{%
  \institution{Huawei Technologies Co., Ltd}
  \city{Shanghai}
  \country{China}
}   
\email{mktb@whut.edu.cn}

\author{Xiaohui Tao}
\affiliation{%
  \institution{University of Southern Queensland}
  \city{Springfield}
  \country{Australia}
}
\email{Xiaohui.Tao@unisq.edu.au}

\author{Jianwei Zhang}
\affiliation{%
  \institution{Iwate University}
  \city{Morioka}
  \country{Japan}
}
\email{zhang@iwate-u.ac.jp}

\renewcommand{\shortauthors}{Trovato et al.}

\begin{abstract}
  In the realm of collaborative filtering recommendation systems, Graph Neural Networks (GNNs) have demonstrated remarkable performance but face significant challenges in deployment on resource-constrained edge devices due to their high embedding parameter requirements and computational costs. 
Using common quantization method directly on node embeddings may overlooks their graph based structure, causing error accumulation during message passing and degrading the quality of quantized embeddings.
  To address this, we propose Graph based Node-Aware Dynamic Quantization training for collaborative filtering (GNAQ), a novel quantization approach that leverages graph structural information to enhance the balance between efficiency and accuracy of GNNs for Top-K recommendation. GNAQ introduces a node-aware dynamic quantization strategy that adapts quantization scales to individual node embeddings by incorporating graph interaction relationships. Specifically, it initializes quantization intervals based on node-wise feature distributions and dynamically refines them through message passing in GNN layers. This approach mitigates information loss caused by fixed quantization scales and captures hierarchical semantic features in user-item interaction graphs. Additionally, GNAQ employs graph relation-aware gradient estimation to replace traditional straight-through estimators, ensuring more accurate gradient propagation during training. Extensive experiments on four real-world datasets demonstrate that GNAQ outperforms state-of-the-art quantization methods, including BiGeaR and N2UQ, 
  by achieving average improvement in 27.8\% Recall@10 and 17.6\% NDCG@10 under 2-bit quantization. In particular, GNAQ is capable of maintaining the performance of full-precision models while reducing their model sizes by 8 to 12 times; in addition, the training time is twice as fast compared to quantization baseline methods.

\end{abstract}





\maketitle

\section{Introduction}
The rapid proliferation of online services and social platforms has led to an explosion of user-item interaction data, making recommender systems indispensable for personalized information filtering~\cite{wu2022survey}. Collaborative filtering (CF) models, especially those based on graph neural networks (GNNs)~\cite{wang2019neural,he2020lightgcn,wu2022graph,yang2021hyper,li2024mealrec+,li2024boosting}, have achieved remarkable success in capturing complex user-item relationships and improving the precision of the recommendation. However, the increasing scale of GNN-based recommender systems poses significant challenges in terms of computational cost and memory consumption, especially when deploying models on resource-constrained devices.

\begin{figure*}[h]
  \centering
  \includegraphics[width=0.9\linewidth]{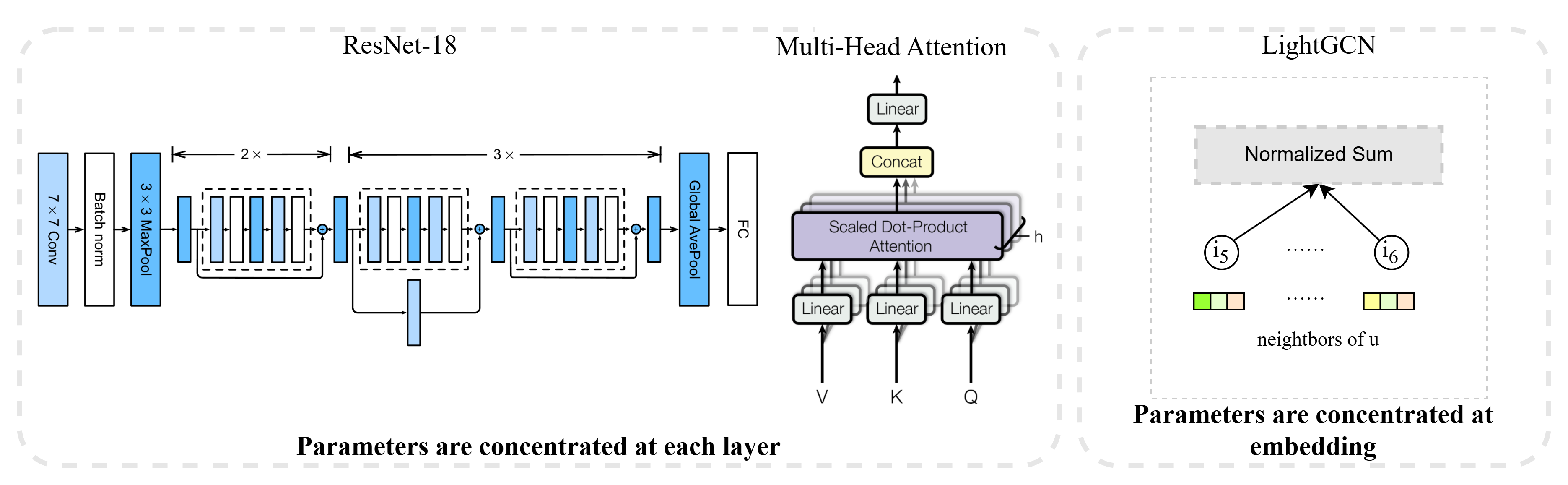}
  \caption{ Embeddings as Parameters of LightCNN in RS.}
  \Description{the request body, tags, and request contributions}
  \label{fig:fig1}
\end{figure*}

Model quantization~\cite{gray1998quantization} has emerged as a promising solution to reduce the storage and computation requirements of deep learning models by representing parameters and activations with low-precision values. Although quantization-aware training (QAT)~\cite{gholami2022survey} and post-training quantization (PTQ)~\cite{xiao2023smoothquant,shao2024omniquant,frantaroptq,lin2024awq} have been extensively studied in computer vision and natural language processing, their application to GNNs for collaborative filtering remains underexplored~\cite{chen2022learning,feng2020sgquant}. The main object of quantization of graph neural networks in recommendation systems is node encoding. As shown in Figure~\ref{fig:fig1}, unlike CNNs, RNNs, or Transformers~\cite{vaswani2017attention}, where the parameters are concentrated mainly in linear, normalization, or attention layers, GNNs for recommendation tasks primarily store the parameters in the embeddings of nodes~\cite{wang2019neural,he2020lightgcn}. During training, node embeddings are optimized via message passing on the graph~\cite{kipf2016semi,wu2020comprehensive}. The final node embeddings serve as representations for users and items, and user-item relevance is computed via dot product or fully connected layers. Thus, quantization in this context essentially refers to quantizing node embeddings~\cite{feng2020sgquant,chen2022learning}.

The distribution of node embeddings is difficult to capture due to their role in fitting interaction relationships~\cite{koren2009matrix,pan2020new}. Directly applying quantization methods from other domains to GNNs in recommendation tasks ignores the unique structure and can lead to error accumulation during message passing, ultimately degrading the quality of quantized embeddings. Moreover, using straight-through estimators~\cite{bengio2013estimating} for gradient approximation may result in parameter updates that are inconsistent with the quantization objective, slowing convergence. Existing QAT methods for GNNs often treat all nodes and parameters uniformly~\cite{liu2022nonuniform}, neglecting the heterogeneity and structural information of graph data. This can lead to suboptimal quantization, as the various roles and connectivity patterns of nodes are not adequately considered~\cite{zhu2023rm}.

To address these limitations, we propose a novel Graph Node-Aware Quantization with dynamic step size (GNAQ) approach for GNN-based collaborative filtering. Our approach introduces three key innovations: (1) node-aware parameter sharing quantization~\cite{liang2024lightweight}, which adapts quantization parameters to the local structure of each node; (2) dynamic step size quantization~\cite{esserlearned,shao2024omniquant}, which learns quantization intervals during training to better fit the distribution of node embeddings; and (3) graph interaction-aware quantization updates, which leverages neighborhood information to refine quantized representations. By integrating these components, GNAQ enables precise and adaptive quantization, leading to improved recommendation performance and efficiency.

Extensive experiments are conducted on four widely used public datasets, demonstrating that our GNAQ consistently outperforms state-of-the-art quantization methods in terms of Recall and NDCG metrics, while maintaining low computational overhead. Our contributions can be summarized as follows. 1) We identify the limitations of existing GNN quantization methods in collaborative filtering and highlight the importance of structure-aware quantization. 2) We propose GNAQ, a novel quantization framework that incorporates node-aware parameter sharing, dynamic step size learning, and graph interaction-aware updates. 3) The comprehensive experimental results show the effectiveness and efficiency of GNAQ in multiple benchmark datasets.


\section{Related Work}

To establish comprehensive technical context for our work, we first review fundamental components spanning three critical dimensions: graph-based collaborative filtering architectures that form our methodological foundation, general model quantization techniques that provide theoretical underpinnings for parameter compression, and specialized quantization approaches for graph neural networks that reveal existing limitations in structural awareness. This tripartite analysis will systematically position our work within the current research landscape.

\subsection{Graph Collaborative Filtering}

Recent advances in graph neural networks have established Graph Collaborative Filtering (GCF) as a dominant paradigm for modeling user-item interactions. Early works like NGCF [26] pioneered the construction of bipartite interaction graphs to capture high-order collaborative signals through multi-layer message passing. Subsequent innovations such as LightGCN [11] demonstrated that simplified architectures with feature transformation removal could achieve superior performance, highlighting the critical role of graph structural learning in recommendation tasks. The emergence of graph contrastive learning techniques further enhanced GCF's capability to handle data sparsity, with methods like SGL~\cite{wu2021self} and HCCF~\cite{meng2025heterogeneous} leveraging self-supervised learning to refine node representations through augmented graph views.

Despite these advancements, conventional GCF architectures face inherent scalability limitations due to their reliance on full-precision node embeddings. The storage complexity grows linearly with the number of entities, requiring gigabyte-level memory for million-scale user/item scenarios. This creates significant deployment barriers for edge devices while neglecting the latent structural patterns that could guide efficient parameter allocation. Existing solutions predominantly focus on architectural modifications rather than fundamentally addressing the precision-redundancy in node representations, leaving the potential of structure-aware quantization largely unexplored.

\begin{table*}[t]
\small
  \caption{Symbol Summary.}
  \label{tab:tab1Symbol}
  \begin{tabular}{cccc}
    \toprule
     Notation   & Description & Notation &  Description \\
     \midrule
    $U$    & Set of users     & $q_{i}$   &  Low-precision embedding representation of entity $i$       \\
$I$    & Set of items      & $a_{i}$   & Scaling factor for the quantization interval      \\
$N$ & Total number of entities   & $e_{i}^{\prime}$ & Extended quantized embedding representation \\
$X$ & User-item interaction matrix   & $E$ & Full-precision embedding table \\
$P$ & Full-precision GNN model with floating-point parameters.   & $E_{\mathrm{quant}}$ & Low-precision embedding table after quantization \\
$Q$ & Quantized GNN model with low-precision parameters.   & $E^{\prime}$ & Extended quantization embedding table \\
$e_{i}$ & Full-precision embedding representation of entity  $i$  & $E_{quant}^{\prime}$ & Embedding table in the forward propagation process \\
     \bottomrule
  \end{tabular}
\end{table*}

\subsection{Model Quantization}

Model quantization aims to reduce the memory footprint and computational cost of deep learning models by representing weights and activations with low-precision values. Quantization-aware training (QAT) and post-training quantization (PTQ) are two mainstream approaches. QAT simulates quantization effects during training, allowing the model to adapt to low-precision constraints, while PTQ applies quantization after model training, often with minimal or no retraining. Although quantization has been extensively studied in computer vision~\cite{courbariaux2016binarized,jacob2018quantization} and NLP~\cite{zafrir2019q8bert}, its application to GNNs, especially for recommender systems, remains limited.

\subsection{Quantization for GNNs}

Recent work has explored quantization techniques for GNNs to enable efficient inference on large-scale graphs. N2UQ introduces non-uniform-to-uniform quantization for GNNs, while BiGeaR proposes a binarized graph representation learning method with multi-faceted quantization reinforcement. However, most existing methods treat all nodes and parameters uniformly, ignoring the heterogeneity of graph structures. Degree-Quant and SGQuant~\cite{feng2020sgquant} attempt to address this by considering node degrees or specialized quantization, but still lack fine-grained adaptation to node-specific characteristics.

Despite recent progress, there remains a gap in structure-aware quantization methods that can dynamically adapt quantization parameters to the local graph structure with respect to node embeddings. Our approach addresses this gap by proposing a node-aware, dynamically adaptive quantization framework for GNN-based collaborative filtering.

\section{Problem Description}
The problem description in the paper is as follows.
We define $U=\{u_1,u_2,\ldots,u_n\}$ and $I=\{i_1,i_2,\ldots,i_m\}$ as the sets of users and items, respectively, where n and m denote the number of users and items. The total number of entities is $N=n+m$. Based on the historical interaction data between users and items, we define the interaction matrix $X=\{x_{ui}|x\in U,i\in I\}$, where $x_{ui}=1$ indicates that user u has interacted with item i, and $x_{ui}=0$ otherwise. P and Q represent the full-precision graph neural network model and the quantized graph neural network model, respectively.

Given the above definitions, the quantization task of graph neural network models in the context of collaborative filtering recommendation can be formally described as follows:

\textbf{Input:} A graph neural network model $P$ with full-precision floating-point parameters; User set $U$; Item set $I$; User-item interaction matrix $X$.
    
\textbf{Output:} A quantized graph neural network model $Q$ with compressed parameters.

To ensure consistency in notation, Table~\ref{tab:tab1Symbol} lists the definitions of high-frequency symbols used in this paper.

\section{GNAQ Methodology}

\begin{figure*}[h]
  \centering
  \includegraphics[width=0.9\linewidth]{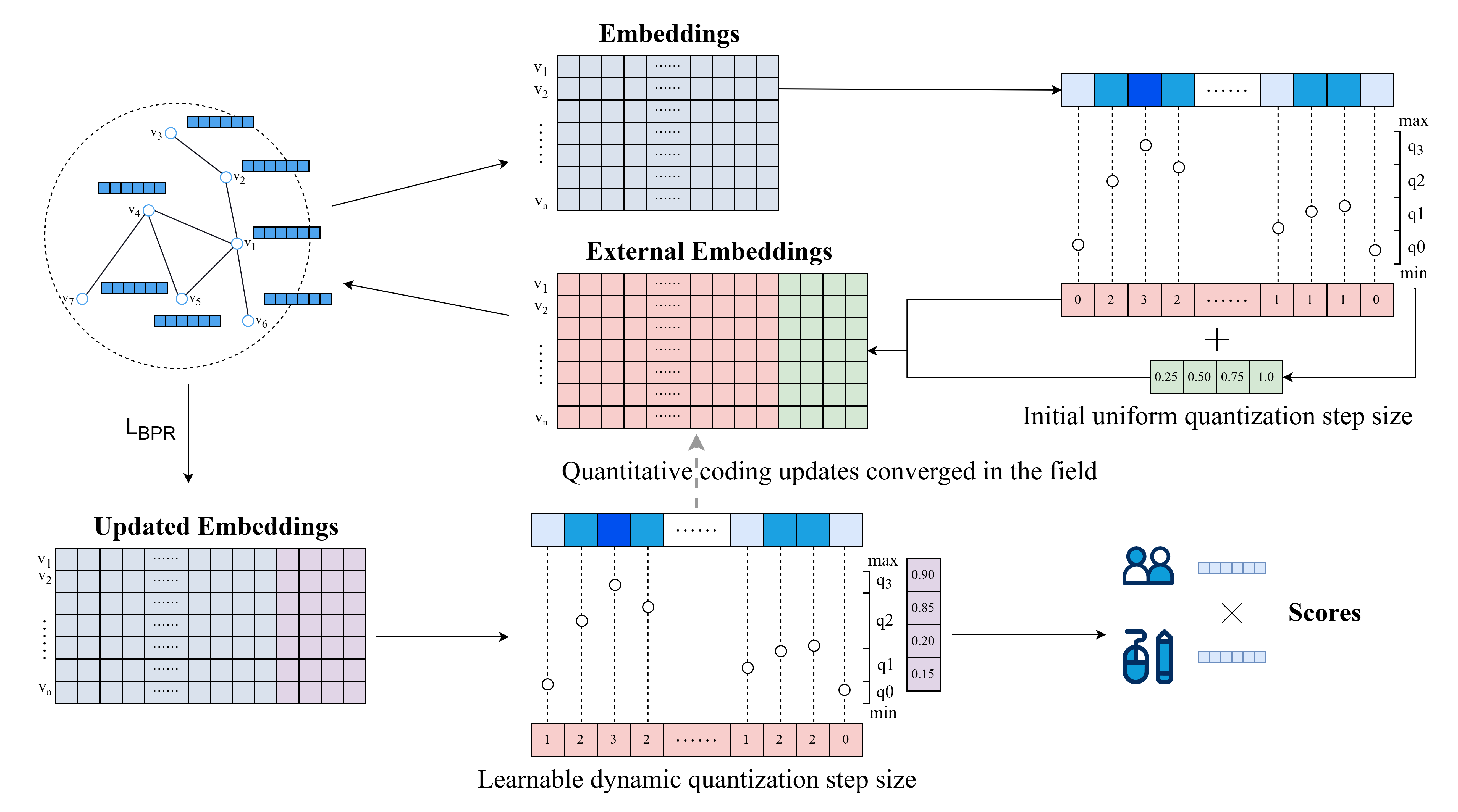}
  \caption{Overall Framework of Our GNAQ.}
  \Description{the request body, tags, and request contributions}
  \label{fig:fig2framework}
\end{figure*}

Considering the structural characteristics of GNNs and the specifics of recommendation scenarios helps to improve the quantization effect of GNN-based recommendation models. Our approach Graph Node-Aware Quantization-aware training (GNAQ) with dynamic step size and asymmetric quantization for collaborative graph filtering recommendation. As shown in Figure~\ref{fig:fig2framework}, this approach is based on quantization-aware training and parameter sharing, transforming the search for quantization parameters into learning of the parameters of the shared model. The optimization of quantization step size is incorporated into the objective of quantized model parameter tuning, enabling effective dynamic adjustment of the step size. Furthermore, a backpropagation method for quantization functions based on graph interaction aggregation is designed, fully considering the mechanism of message passing of GNNs. Finally, to address Top N rank recommendation scenarios, a ranking loss is added to the quantization training loss function to enhance the ranking ability of the quantized model. In summary, the main features of our proposed GNAQ are as follows.

\textbf{Parameter sharing for quantization-aware training:} GNAQ leverages the rich node association information in GNNs and implements quantization-aware training based on parameter sharing.

\textbf{Dynamic step size update via message passing:} The quantization step size is concatenated with the embeddings of the nodes, and dynamic updates are achieved through the GNN message passing mechanism. The quantized encoding is updated via graph interaction aggregation, effectively updating quantized values.

\textbf{Ranking loss in quantization training:} The loss function includes a ranking loss, which effectively improves the ranking ability of the quantized model.


\subsection{Quantization Based on Parameter Sharing}

Parameter sharing is a method to reduce the total number of parameters by reusing the same parameters in different parts of a neural network model. This method defines an allocation matrix to associate the encoding of each node with the encoding in the meta-embedding. During training, the encoding representation of each node is composed by combining the meta-embedding according to the allocation matrix. Through the backpropagation mechanism of the graph neural network, the update of the meta-embedding parameters is realized. Similarly, if the allocation matrix is used as the quantization encoding representation of the node and the meta-embedding is used as the quantization step size of each node, the quantization step size of the node can be determined using the backpropagation of the graph neural network through the idea of parameter sharing.

Our approach realizes graph node-aware dynamic quantization step size asymmetric $n$-bit quantization by assigning $2n$ quantization step sizes to each node. Using the idea of parameter sharing, the quantization step size of each node is concatenated into the quantization encoding of the node, and through the information transfer of the graph neural network, the dynamic update of the node quantization encoding and the quantization step size is achieved. Unlike traditional quantization methods that set fixed quantization functions to quantize parameters within a fixed range to specified quantization values, GNAQ deals with parameter quantization by using an allocation matrix similar to that in parameter sharing methods. During the learning process of the model for the allocation matrix and the quantization step size, the quantization function is learned dynamically, and the quantization of the graph neural network is carried out efficiently.

\subsection{Initialization of Quantization Step Size}

Our approach uses the node encoding table $E_{full}\in\mathbb{R}^{N\times d}$ of the full-precision model as the initial values of the quantized model. Then, the maximum and minimum values in each node encoding are obtained. According to the maximum and minimum values of each node embedding, the initial quantization step size $gap_i$ of each node encoding is calculated as:

\begin{align}
gap_i & = \frac{\max(e_i)-\min(e_i)}{2^b}\label{gap},
\end{align}
where $e_{i}$ is the full-precision encoding of node $i$; $N$ represents the total number of nodes; $n$ represents the quantization precision, that is, the bit width used for quantization; $max$ represents the maximum value by row; $min$ is the minimum value by row.

The initial quantization step size is uniform. Based on the maximum and minimum values of each node encoding and the quantization bit width b, the value range of each node encoding is evenly divided into $2^{n}$ intervals. The initial quantization function based on these intervals is:

\begin{align} \label{eqa2}
Q_i(a) = \begin{cases}
0, & \min_i \leq a < \min_i + gap_i \\
1, & \min_i + gap_i \leq a < \min_i + 2*gap_i \\
\vdots \\
2^n-1, & \min_i + (2^n-1)*\mathrm{gap}_i \leq a \leq \max_i
\end{cases}
\end{align}

The above formula represents the quantization function for quantizing the embedding representation $e_i\in\mathbb{R}^d$ of the $i\mathrm{-th}$  node. $\mathrm{min}_i$ is the minimum value in $e_{i}$, $\mathrm{max}_i$ is the maximum value in $e_{i}$, and $gap_i$ is the quantization interval obtained by calculating according to Equation~\ref{gap}. The quantization function corresponding to each node encoding is executed to obtain the quantization encoding $q_{i}$ and $E_{quant}$ of each node:

\begin{align}
q_i & = [Q_i(e_i[0]),Q_i(e_i[1]),\cdots\cdots,Q_i(e_i[d-1])]
\end{align}

\begin{align}
E_{quant} & = [q_0,q_1,\cdots\cdots,q_{N-1}]^T
\end{align}

\subsection{Update of Quantization Interval Scaling}

The midpoint value of each quantization interval is used as the scaling factor of the corresponding node encoding. The scaling factor of the quantization interval for a node is computed as:

\begin{align} \label{eqa5}
s_i & = \left[{\min}_i+\frac{\mathrm{gap}_i}2,{\min}_i+\frac{3*\mathrm{gap}_i}2,\cdots\cdots,{\min}_i+\frac{(2^{n+1}-1)*\mathrm{gap}_i}2\right]
\end{align}

\begin{align}
S & = [s_0,s_1,\cdots,s_{N-1}]^\mathrm{T}
\end{align}

where $a_i\in\mathbb{R}^{2^n}$. The scaling factor of the quantization interval is represented by a low-precision floating-point number (FP8~\cite{micikevicius2022fp8}), which ensures the efficiency of subsequent calculations. Then, the scaling factor of the quantization interval, defined in Equations~\ref{eqa7} to ~\ref{eqa8}, is concatenated into the node encoding to form the extended quantization encoding table E. 

\begin{align} \label{eqa7}
e_i^{\prime} & = [q_i,s_i]
\end{align}

\begin{align}\label{eqa8}
E & = [e_0^{\prime},e_1^{\prime},\cdots,e_{n-1}^{\prime}],
\end{align}
where $e_i^{\prime}\in\mathbb{R}^{d+2^n}$, $E\in\mathbb{R}^N\times(d+2^n)$. Using the extended quantization encoding, the quantization interval scaling factor can be updated with the help of the message-passing mechanism of the graph neural network, thus achieving dynamic quantization step sizes. In this way, different quantization functions can be constructed for each node encoding, improving the quantization's perception of node information and thus enhancing the quantization effect.

After completing the update of the scaling factor of the quantization interval, it is necessary to dynamically adjust the quantization step size. Therefore, the dynamic update mechanism of the quantization step size is described in detail below.

\subsection{Dynamic Quantization Step Size}\label{4.4}

\textbf{Forward Propagation:} 
During the forward propagation process of the graph neural network, our extended quantization encoding is used as the input, and message passing is performed on the graph. Before inputting the extended quantization embedding into the graph neural network, it is necessary to transform the quantization embedding $E_{quant}$, mapping the quantization encoding to the quantization interval scaling factor:

\begin{align}
q_i^{\prime} & = \left[s_i[q_i[0]],s_i[q_i[1]],\cdots\cdots,s_i[q_i[d-1]]\right]
\end{align}

\begin{align}
E_{\mathrm{quant}}^{\prime} & = [q_0^{\prime},q_1^{\prime},\cdots\cdots,q_{N-1}^{\prime}]
\end{align}

In this way, the effect of non-uniform quantization is achieved. The transformed quantization encoding is concatenated with the quantization interval scaling factor to obtain the input encoding table $H^{(0)}\in\mathbb{R}^{N\times(d+2^n)}$ of the graph neural network in Equation~\ref{equ11}.

\begin{align} \label{equ11}
H^{(0)} & = [E_{\mathrm{quant}}^{\prime},S]
\end{align}

Then, the input encoding table is used as the initial encoding of the nodes and input into the graph neural network. Through the information transfer of the graph neural network, message passing is performed on the normalized Laplacian matrix to optimize the quantization encoding representation of the nodes and the quantization interval scaling factor. Then, the dynamic quantization step size is achieved through the relationship between the quantization interval scaling factor and the quantization step size.

It can be seen from Equation~\ref{eqa5} and~\ref{eqa7} that the initial quantization clipping interval of GNAQ is uniform. During the subsequent model training process, the quantization clipping interval will also change by learning the quantization interval scaling factor, thus realizing a learnable quantization clipping interval. Forward propagation across GNN layers adheres to:

\begin{align}
H^{(l+1)} & = \left(D^{-\frac12}XD^{-\frac12}\right)H^{(l)},
\end{align}

\begin{align}
H=\frac1{l+1}\sum_{l=0}^LH^{(l)},
\end{align}
where $X$ is the user-item interaction data, represented as an adjacency matrix; $L$ is the number of layers of the graph neural network; D is the degree matrix of $X$, and $D^{-\frac12}XD^{-\frac12}$  forms a symmetric degree-normalized adjacency matrix; the final node encoding representation $H$ is the average value of each layer's representation.

\textbf{Update of Quantization Step Size:}
Since the initial quantization interval scaling factor is added to the node encoding of the graph neural network and calculated together with the encoding during the forward propagation of the graph neural network, the quantization interval scaling factor will be updated during the backpropagation process of the model. Suppose the updated quantization interval scaling factor is $S^{\prime}$. The updated quantization step size can be calculated through $S^{\prime}$ and the current value range of the node encoding. The update method is as follows: First, sort the updated quantization interval scaling factors according to their numerical sizes. Then, take the midpoint positions between adjacent quantization interval scaling factors and the boundaries of the value distribution range as the boundaries of the quantization intervals. There are a total of $2^n-1$ midpoint values, which can divide the value range of the node encoding into $2^n$ intervals, and the length of each interval is the updated quantization step size.

\begin{figure}[h]
  \centering
  \includegraphics[width=0.8\linewidth]{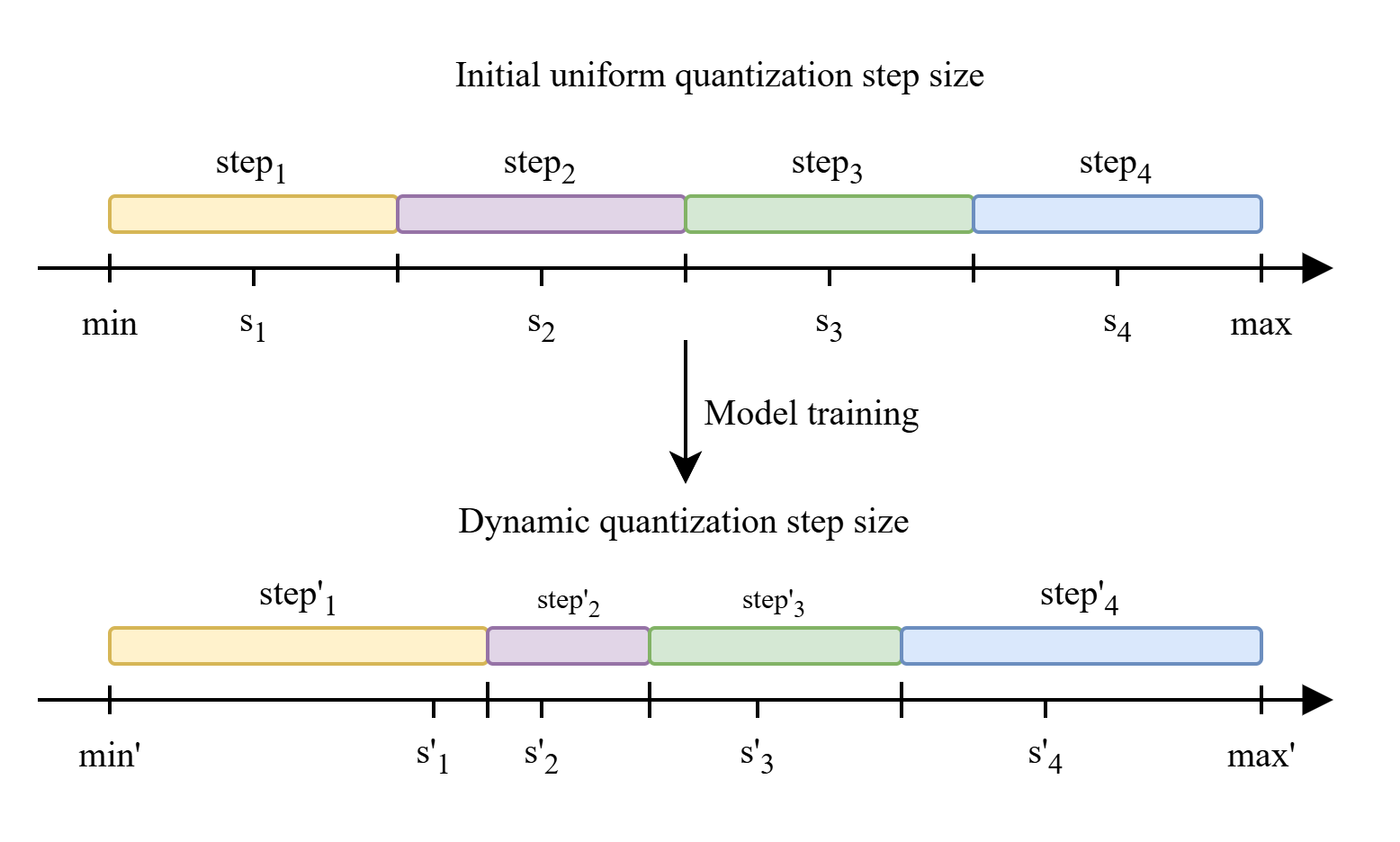}  \caption{Dynamic Update of Quantization Step Size.}
  \Description{the request body, tags, and request contributions}
  \label{fig:fig3dynamic}
\end{figure}

As shown in Figure~\ref{fig:fig3dynamic}, taking 2-bit quantization as an example, each node encoding has four quantization interval scaling factors $s_1,s_2,s_3,\text{and }s_4$. Initially, the intervals of the quantization interval scaling factors are uniform, and the quantization step sizes are also equal (i.e., $\text{step}_1=\text{step}_2=\text{step}_3=\text{step}_4$). After model training, the quantization interval scaling factors are updated. First, rearrange the updated quantization interval scaling factors according to their numerical sizes, such as $s_1^{\prime},s_2^{\prime},s_3^{\prime},\text{and }s_4^{\prime}$ in Figure~\ref{fig:fig3dynamic}. Then, the midpoint of two scaling factors is taken as the new boundary of the quantization interval and four new quantization intervals are obtained. The length of each interval is the updated quantization step size that are no longer equal. In this way, the value of the quantization step size is learned through model training, and the parameter optimization ability of the neural network model is utilized to efficiently optimize the quantization step size.

\textbf{Update of Quantization Encoding Aggregated by Graph Interaction Relationships:}
As shown in Equation~\ref{eqa2}, the quantization function of node encoding is a discrete piecewise function. Therefore, at the edges of the piecewise intervals, the quantization function is non-differentiable; within the piecewise intervals, the derivative is constantly 0. This makes it impossible to update the model parameters through backpropagation during the model training process. Traditional quantization methods approximate the gradient of the quantization function through gradient estimation methods such as the straight through estimator to achieve parameter updates. Since our approach performs model quantization based on the idea of parameter sharing, the node quantization encoding is not directly updated during the backpropagation process of the model. Therefore, this paper combines the ideas of parameter sharing and low precision training (LTP~\cite{li2024embedding}), regards the update of the node quantization encoding as the update of the allocation matrix. After the backpropagation of the model, the information of adjacent nodes is aggregated to achieve the update of the node quantization encoding, thus avoiding the estimation of the gradient of the quantization function.

Specifically, after the backpropagation of the model, the encoding and quantization interval scaling factor of each node will be updated. As described in Section~\ref{4.4}, the quantization step size can be calculated based on the updated quantization interval scaling factor, and then the updated quantization function can be obtained. In GNAQ, the first-order aggregated information of nodes is used as the basis for updating the node quantization encoding, and the calculation formula for the first-order aggregated information of nodes is as follows:

\begin{align}
\hat{e}_i & = \frac1{|\mathcal{N}_i|}\sum_{j\in\mathcal{N}_i}e_j,
\end{align}
where $\mathcal{N}_{i}$ represents the set of neighbor nodes of node i, and $|\mathcal{N}_i|$ represents the number of neighbor nodes of node i. Then, the first-order aggregated information is input into the updated quantization function to obtain the updated node quantization encoding. Using the first-order aggregated information of nodes as the basis for updating the node quantization encoding can avoid the "Z-shaped" oscillation phenomenon~\cite{ma2024one} during the parameter update process. Using the node representation aggregated by graph interaction relationships can limit the model to consider the interaction information in the graph during the update of the node quantization encoding, playing a part of the regularization effect.

\subsection{Loss Function}\

To enhance the quantized model's ability to predict and rank interaction relationships, GNAQ incorporates the Bayesian Personalized Ranking Loss (BPR~\cite{rendle2012bpr}) and LambdaLoss~\cite{wang2018lambdaloss} to optimize the model parameters. The BPR loss is formulated as:

\begin{align}
\mathcal{L}_{\mathrm{BPR}} & = \sum_{(u,i^+,i^-)\in\mathcal{B}}-\ln\sigma(\hat{y}_{ui^+}-\hat{y}_{ui^-})+\lambda\parallel\Theta\parallel^2
\end{align}

\begin{align}
\mathcal{L}_{\mathrm{Lambda}}=-\sum_{i=1}^K\sum_{j=1}^K\eta_{ij}\cdot\log(1+e^{-(s_i-s_j)})
\end{align}

\begin{align}
\mathcal{L} & = \mathcal{L}_{\mathrm{BPR}}+\mathcal{L}_{\mathrm{Lambda}},
\end{align}
where $B$ represents the set of data in each batch; $(u,i^+,i^-)$ is the triple representation of training data, with $u$ denoting the user, $i^{+}$ representing the item that has an explicit interaction with the the user $u$ and $i^{-}$ representing the item that has no explicit interaction with the user. $\parallel\Theta\parallel^2$ is the regularization term, and $\text{λ}$ is the hyperparameter. $\eta_{ij}$ represents the ranking scores of the i-th and j-th items in the prediction sequence. If the  i-th item should be ranked before the j-th item, then $\eta_{ij}$=1; conversely, if the i-th item should be ranked after the j-th item, then $\eta_{ij}$=−1. $s_{i}$ and $s_{j}$ represent the predicted scores of the i-th and j-th items in the prediction sequence, respectively. The BPR loss can optimize the relative preference relationship and effectively handle implicit feedback data. LambdaLoss can directly optimize the ranking metric (NDCG) and re-rank the initially predicted sequence of the model. By combining the BPR loss and LambdaLoss, the model's ability to capture ranking information can be strengthened, effectively improving the model's recommendation performance.

\section{Experiments}
Our approach is evaluated on the Top-K recommendation task with the aim of answering the following research questions:
\begin{itemize}
    \item \textbf{RQ1.} How does GNAQ perform compared to state-of-the-art
full-precision and quantization-based models?
    \item \textbf{RQ2.} How is the practical resource consumption of GNAQ?
    \item \textbf{RQ3.} How do the proposed key components affect GNAQ performance?
\end{itemize}

\begin{table}[htp]
\scriptsize
  \caption{Parameter settings.}
  \label{tab:tab3Parameter}
  \begin{tabular}{ccccc}
    \toprule
     Parameter            & MovieLens          & Gowalla            & Yelp2020           & Amazon-Book        \\
     \midrule
    batch size    & 10240              & 5096               & 5096               & 5096               \\
learning rate & $1\times10^{-3}$            & $1\times10^{-3}$            & $5\times10^{-4}$            & $5\times10^{-4}$           \\
embedding dim & {[}64, 128, 256{]} & {[}64, 128, 256{]} & {[}64, 128, 256{]} & {[}64, 128, 256{]} \\
optimizer     & Adam               & Adam               & Adam               & Adam               \\
GNN layers    & 3                  & 3                  & 3                  & 3                  \\
      ${\lambda}$        & $5\times10^{-4}$             & $5\times10^{-4}$             & $5\times10^{-4}$             & $5\times10^{-4}$  \\          
     \bottomrule
  \end{tabular}
\end{table}

\begin{figure*}[h]
  \centering
  \includegraphics[width=0.85\linewidth]{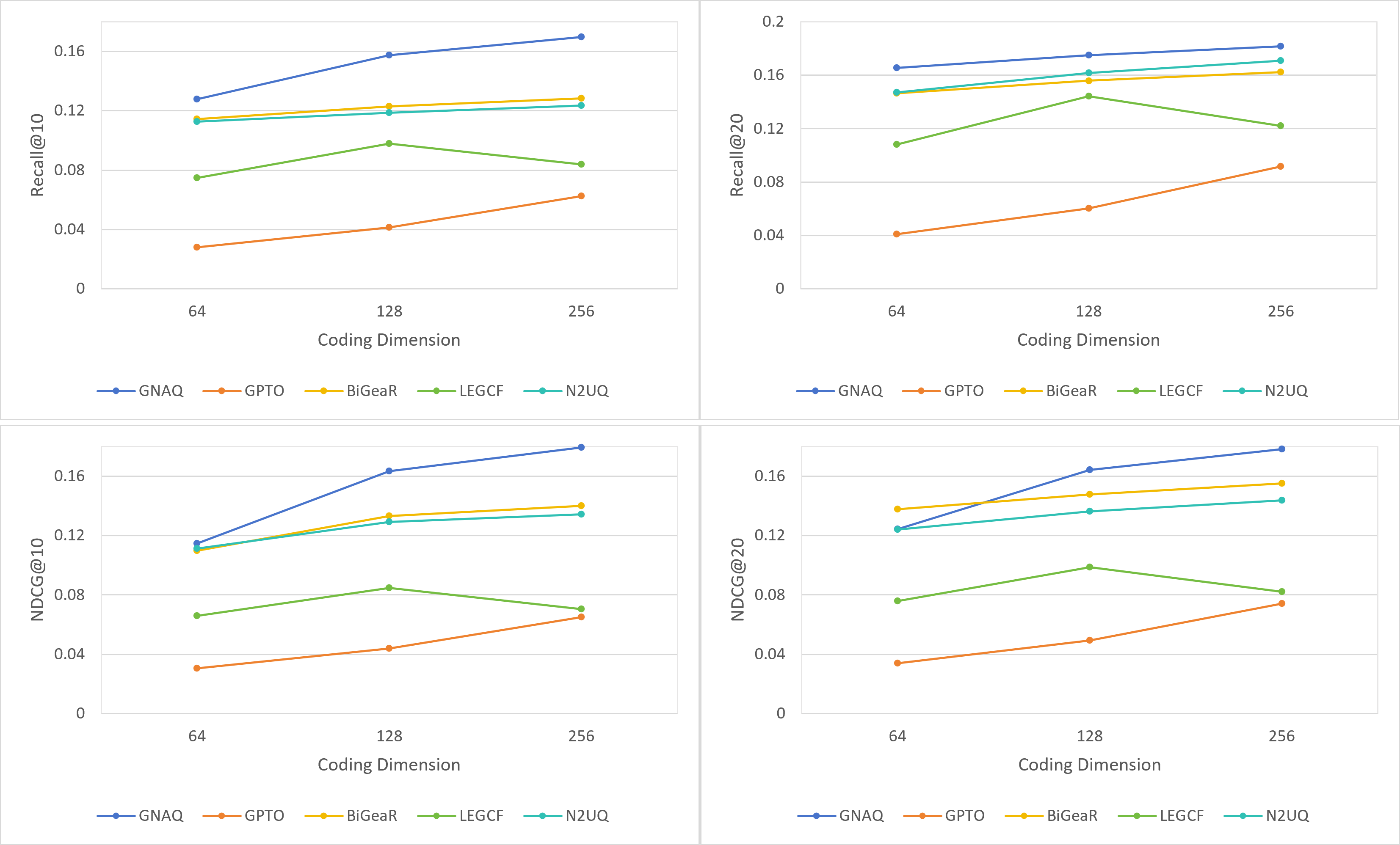}
  \caption{The Effect of Coding Dimension on Recommendation Performance.}
  \Description{the request body, tags, and request contributions}
  \label{fig:fig4Gowalla}
\end{figure*}

\subsection{Experimental Setup}

\textbf{Datasets}. Our evaluation is conducted on the following four datasets:
MovieLens~\cite{harper2015movielens}, Gowalla~\cite{cho2011friendship}, Yelp2020 and Amazon-Book, among which Yelp2020 and Amazon-Book can be found in the official code repositories of LightGCN~\cite{he2020lightgcn} and LEGCF~\cite{liang2024lightweight}.

\textbf{Evaluation Metrics}. Recall@K and NDCG@K are adopted as evaluation metrics, which are standard in the top-K recommendation tasks. Recall@K measures the proportion of relevant items retrieved in the top-K recommendations, while NDCG@K evaluates the ranking quality by considering the positions of relevant items.

\textbf{Baselines.} We compare our GNAQ against following baselines:
\textbf{BiGeaR~\cite{chen2022learning}:} presented by Chen et al. in 2022, it is a multi-faceted quantization-enhanced binarized graph representation learning method. 
\textbf{N2UQ~\cite{liu2022nonuniform}:} a quantization method proposed by Liu et al. in 2022, which transforms non-uniform quantization into uniform quantization and uses Generalized Straight. 
\textbf{GPTQ~\cite{frantaroptq}:} proposed by Frantar et al. in 2023, it is a post-training quantization method that uses layer-wise quantization and OBQ (Optimal Brain Quantization) technique to minimize quantization errors. 
\textbf{LEGCF~\cite{liang2024lightweight}:} put forward by Liang et al. in 2024, it is a lightweight GNN node embedding learning method based on parameter sharing. 

\textbf{Parameter Settings.} The experimental results of all baseline models were reproduced in this paper, following the optimal settings provided in their publications. The learning rate is from the collection
{1e-5,5e-5,1e-4,5e-4,1e-3}, the optimizer selects Adam. In our model, the quantization bitwidth is set to 2 bits. To explore the impact of different node embedding dimensions on the quantization performance of graph neural network models, experiments were conducted under three embedding dimensions (64-dimensional, 128-dimensional, and 256-dimensional). Parameters such as batch size and training epochs were adjusted according to the node count and interaction data of different datasets, as detailed in Table~\ref{tab:tab3Parameter}.

We employ LightGCN~\cite{he2020lightgcn}, the current state-of-the-art graph neural network model for collaborative filtering tasks, as the baseline model. During model quantization, the initial parameters loaded are those of the pre-trained LightGCN model. Our source codes are available at:https://github.com/WUT-IDEA.

\begin{table*}[htp]

  \caption{Results of Recall@ and NDCG@ for GNAQ vs. Baseline Models with 64-dimensional Embeddings.}
  \label{tab:tab4dimensional64}
  \begin{tabular}{cccccccccc}
    \hline
\multirow{2}{*}{Model} &
   &
  \multicolumn{2}{c}{MovieLens} &
  \multicolumn{2}{c}{Gowalla} &
  \multicolumn{2}{c}{Yelp2020} &
  \multicolumn{2}{c}{Amazon-Book} \\ \cline{3-10} 
 &
  Metric &
  Recall &
  NDCG &
  Recall &
  NDCG &
  Recall &
  NDCG &
  Recall &
  NDCG \\ \hline
  
LightGCN~\cite{he2020lightgcn} &
  \begin{tabular}[c]{@{}c@{}}@10\\ @20\end{tabular} &
  \begin{tabular}[c]{@{}c@{}}0.1551\\ 0.2418\end{tabular} &
  \begin{tabular}[c]{@{}c@{}}0.2742\\ 0.2801\end{tabular} &
  \begin{tabular}[c]{@{}c@{}}0.1088\\ 0.1557\end{tabular} &
  \begin{tabular}[c]{@{}c@{}}0.0884\\ 0.1041\end{tabular} &
  \begin{tabular}[c]{@{}c@{}}0.0409\\ 0.0687\end{tabular} &
  \begin{tabular}[c]{@{}c@{}}0.0274\\ 0.0366\end{tabular} &
  \begin{tabular}[c]{@{}c@{}}0.0204\\ 0.0353\end{tabular} &
  \begin{tabular}[c]{@{}c@{}}0.0165\\ 0.0222\end{tabular} \\
  \hline
BiGeaR~\cite{chen2022learning} &
  \begin{tabular}[c]{@{}c@{}}@10\\ @20\end{tabular} &
      \begin{tabular}[c]{@{}c@{}}\underline{0.1479}\\ \underline{0.1796}\end{tabular} &
  \begin{tabular}[c]{@{}c@{}}\underline{0.3723}\\ \underline{0.3554}\end{tabular} &
  \begin{tabular}[c]{@{}c@{}}\underline{0.1144}\\ 0.1466\end{tabular} &
  \begin{tabular}[c]{@{}c@{}}0.1098\\ \textbf{0.1378}\end{tabular} &
  \begin{tabular}[c]{@{}c@{}}\underline{0.0452}\\ \textbf{0.0712}\end{tabular} &
  \begin{tabular}[c]{@{}c@{}}\underline{0.0393}\\ \textbf{0.0498}\end{tabular} &
  \begin{tabular}[c]{@{}c@{}}\textbf{0.0187}\\ 0.0266\end{tabular} &
  \begin{tabular}[c]{@{}c@{}}\underline{0.0161}\\ 0.0206\end{tabular} \\
N2UQ~\cite{liu2022nonuniform} &
  \begin{tabular}[c]{@{}c@{}}@10\\ @20\end{tabular} &
  \begin{tabular}[c]{@{}c@{}}0.1093\\ 0.1692\end{tabular} &
  \begin{tabular}[c]{@{}c@{}}0.2494\\ 0.2419\end{tabular} &
  \begin{tabular}[c]{@{}c@{}}0.1127\\ \underline{0.1471}\end{tabular} &
  \begin{tabular}[c]{@{}c@{}}\underline{0.1113}\\ 0.1204\end{tabular} &
  \begin{tabular}[c]{@{}c@{}}0.0324\\ 0.0520\end{tabular} &
  \begin{tabular}[c]{@{}c@{}}0.0371\\ 0.0418\end{tabular} &
  \begin{tabular}[c]{@{}c@{}}0.0157\\ \underline{0.0282}\end{tabular} &
  \begin{tabular}[c]{@{}c@{}}\textbf{0.0168}\\ \textbf{0.0218}\end{tabular} \\
GPTQ~\cite{frantaroptq} &
  \begin{tabular}[c]{@{}c@{}}@10\\ @20\end{tabular} &
  \begin{tabular}[c]{@{}c@{}}0.0653\\ 0.1066\end{tabular} &
  \begin{tabular}[c]{@{}c@{}}0.0897\\ 0.1027\end{tabular} &
  \begin{tabular}[c]{@{}c@{}}0.0280\\ 0.0411\end{tabular} &
  \begin{tabular}[c]{@{}c@{}}0.0304\\ 0.0340\end{tabular} &
  \begin{tabular}[c]{@{}c@{}}0.0106\\ 0.0180\end{tabular} &
  \begin{tabular}[c]{@{}c@{}}0.0091\\ 0.0117\end{tabular} &
  \begin{tabular}[c]{@{}c@{}}0.0045\\ 0.0079\end{tabular} &
  \begin{tabular}[c]{@{}c@{}}0.0052\\ 0.0064\end{tabular} \\
LEGCF~\cite{liang2024lightweight} &
  \begin{tabular}[c]{@{}c@{}}@10\\ @20\end{tabular} &
  \begin{tabular}[c]{@{}c@{}}0.0705\\ 0.1160\end{tabular} &
  \begin{tabular}[c]{@{}c@{}}0.1435\\ 0.1427\end{tabular} &
  \begin{tabular}[c]{@{}c@{}}0.0748\\ 0.1082\end{tabular} &
  \begin{tabular}[c]{@{}c@{}}0.0660\\ 0.0759\end{tabular} &
  \begin{tabular}[c]{@{}c@{}}0.0223\\ 0.0400\end{tabular} &
  \begin{tabular}[c]{@{}c@{}}0.0160\\ 0.0218\end{tabular} &
  \begin{tabular}[c]{@{}c@{}}0.0115\\ 0.0204\end{tabular} &
  \begin{tabular}[c]{@{}c@{}}0.0100\\ 0.0133\end{tabular} \\
\textbf{GNAQ (Ours)} &
  \begin{tabular}[c]{@{}c@{}}@10\\ @20\end{tabular} &
  \textbf{\begin{tabular}[c]{@{}c@{}}0.2048\\ 0.2055\end{tabular}} &
  \textbf{\begin{tabular}[c]{@{}c@{}}0.4934\\ 0.3902\end{tabular}} &
  \textbf{\begin{tabular}[c]{@{}c@{}}0.1277\\ 0.1605\end{tabular}} &
  \begin{tabular}[c]{@{}c@{}}\textbf{0.1145}\\ \underline{0.1243}\end{tabular} &
  \begin{tabular}[c]{@{}c@{}}\textbf{0.0540}\\ \underline{0.0685}\end{tabular} &
  \begin{tabular}[c]{@{}c@{}}\textbf{0.0428}\\ \underline{0.0470}\end{tabular} &
  \begin{tabular}[c]{@{}c@{}}\underline{0.0186}\\ \textbf{0.0319}\end{tabular} &
  \begin{tabular}[c]{@{}c@{}}0.0157\\ \underline{0.0206}\end{tabular} \\
  \hline
\%Improv. &
  \begin{tabular}[c]{@{}c@{}}@10\\ @20\end{tabular} &
  \begin{tabular}[c]{@{}c@{}}38.4\%\\ 14.4\%\end{tabular} &
  \begin{tabular}[c]{@{}c@{}}12.1\%\\ 9.79\%\end{tabular} &
  \begin{tabular}[c]{@{}c@{}}11.6\%\\ 12.5\%\end{tabular} &
  \begin{tabular}[c]{@{}c@{}}2.88\%\\ -9.80\%\end{tabular} &
  \begin{tabular}[c]{@{}c@{}}19.4\%\\ -3.79\%\end{tabular} &
  \begin{tabular}[c]{@{}c@{}}8.91\%\\ -5.62\%\end{tabular} &
  \begin{tabular}[c]{@{}c@{}}-0.53\%\\ 13.1\%\end{tabular} &
  \begin{tabular}[c]{@{}c@{}}-6.55\%\\ -5.50\%\end{tabular}\\ \hline
  \end{tabular}
\end{table*}

\begin{table*}[t]

  \caption{Results of Recall@ and NDCG@ for GNAQ vs. Baseline Models with 128-dimensional Embeddings.}
  \label{tab:tab5dimensional128}
  \begin{tabular}{cccccccccc}
    \hline
\multirow{2}{*}{Model} &
   &
  \multicolumn{2}{c}{MovieLens} &
  \multicolumn{2}{c}{Gowalla} &
  \multicolumn{2}{c}{Yelp2020} &
  \multicolumn{2}{c}{Amazon-Book} \\ \cline{3-10} 
 &
  Metric &
  Recall &
  NDCG &
  Recall &
  NDCG &
  Recall &
  NDCG &
  Recall &
  NDCG \\ \hline

LightGCN~\cite{he2020lightgcn} &
  \begin{tabular}[c]{@{}c@{}}@10\\ @20\end{tabular} &
  \begin{tabular}[c]{@{}c@{}}0.1658\\ 0.2585\end{tabular} &
  \begin{tabular}[c]{@{}c@{}}0.2855\\ 0.2935\end{tabular} &
  \begin{tabular}[c]{@{}c@{}}0.1101\\ 0.1576\end{tabular} &
  \begin{tabular}[c]{@{}c@{}}0.0901\\ 0.1059\end{tabular} &
  \begin{tabular}[c]{@{}c@{}}0.0426\\ 0.0721\end{tabular} &
  \begin{tabular}[c]{@{}c@{}}0.0284\\ 0.0382\end{tabular} &
  \begin{tabular}[c]{@{}c@{}}0.0215\\ 0.0367\end{tabular} &
  \begin{tabular}[c]{@{}c@{}}0.0172\\ 0.0230\end{tabular} \\
  \hline
BiGeaR~\cite{chen2022learning} &
  \begin{tabular}[c]{@{}c@{}}@10\\ @20\end{tabular} &
  \begin{tabular}[c]{@{}c@{}}\underline{0.1551}\\ 0.1788\end{tabular} &
  \begin{tabular}[c]{@{}c@{}}\underline{0.3894}\\ \underline{0.3740}\end{tabular} &
  \begin{tabular}[c]{@{}c@{}}\underline{0.1229}\\ 0.1558\end{tabular} &
  \begin{tabular}[c]{@{}c@{}}\underline{0.1331}\\ \underline{0.1478}\end{tabular} &
  \begin{tabular}[c]{@{}c@{}}\underline{0.0503}\\ \underline{0.0762}\end{tabular} &
  \begin{tabular}[c]{@{}c@{}}\underline{0.0433}\\ \underline{0.0548}\end{tabular} &
  \begin{tabular}[c]{@{}c@{}}\underline{0.0219}\\ \underline{0.0379}\end{tabular} &
  \begin{tabular}[c]{@{}c@{}}0.0177\\ 0.0228\end{tabular} \\
N2UQ~\cite{liu2022nonuniform} &
  \begin{tabular}[c]{@{}c@{}}@10\\ @20\end{tabular} &
  \begin{tabular}[c]{@{}c@{}}0.1272\\ \underline{0.1861}\end{tabular} &
  \begin{tabular}[c]{@{}c@{}}0.2889\\ 0.2569\end{tabular} &
  \begin{tabular}[c]{@{}c@{}}0.1188\\ \underline{0.1618}\end{tabular} &
  \begin{tabular}[c]{@{}c@{}}0.1291\\ 0.1363\end{tabular} &
  \begin{tabular}[c]{@{}c@{}}0.0350\\ 0.0564\end{tabular} &
  \begin{tabular}[c]{@{}c@{}}0.0399\\ 0.0459\end{tabular} &
  \begin{tabular}[c]{@{}c@{}}0.0188\\ 0.0334\end{tabular} &
  \begin{tabular}[c]{@{}c@{}}\underline{0.0198}\\ \underline{0.0257}\end{tabular} \\
GPTQ~\cite{frantaroptq} &
  \begin{tabular}[c]{@{}c@{}}@10\\ @20\end{tabular} &
  \begin{tabular}[c]{@{}c@{}}0.0867\\ 0.1398\end{tabular} &
  \begin{tabular}[c]{@{}c@{}}0.0949\\ 0.1153\end{tabular} &
  \begin{tabular}[c]{@{}c@{}}0.0414\\ 0.0603\end{tabular} &
  \begin{tabular}[c]{@{}c@{}}0.0440\\ 0.0494\end{tabular} &
  \begin{tabular}[c]{@{}c@{}}0.0131\\ 0.0230\end{tabular} &
  \begin{tabular}[c]{@{}c@{}}0.0111\\ 0.0146\end{tabular} &
  \begin{tabular}[c]{@{}c@{}}0.0067\\ 0.0119\end{tabular} &
  \begin{tabular}[c]{@{}c@{}}0.0076\\ 0.0095\end{tabular} \\
LEGCF~\cite{liang2024lightweight} &
  \begin{tabular}[c]{@{}c@{}}@10\\ @20\end{tabular} &
  \begin{tabular}[c]{@{}c@{}}0.0737\\ 0.1241\end{tabular} &
  \begin{tabular}[c]{@{}c@{}}0.1523\\ 0.1525\end{tabular} &
  \begin{tabular}[c]{@{}c@{}}0.0979\\ 0.1444\end{tabular} &
  \begin{tabular}[c]{@{}c@{}}0.0846\\ 0.0988\end{tabular} &
  \begin{tabular}[c]{@{}c@{}}0.0310\\ 0.0548\end{tabular} &
  \begin{tabular}[c]{@{}c@{}}0.0214\\ 0.0291\end{tabular} &
  \begin{tabular}[c]{@{}c@{}}0.0156\\ 0.0259\end{tabular} &
  \begin{tabular}[c]{@{}c@{}}0.0134\\ 0.0172\end{tabular} \\
\textbf{GNAQ (Ours)} &
  \begin{tabular}[c]{@{}c@{}}@10\\ @20\end{tabular} &
  \textbf{\begin{tabular}[c]{@{}c@{}}0.2287\\ 0.2296\end{tabular}} &
  \textbf{\begin{tabular}[c]{@{}c@{}}0.5261\\ 0.4194\end{tabular}} &
  \textbf{\begin{tabular}[c]{@{}c@{}}0.1576\\ 0.1750\end{tabular}} &
  \textbf{\begin{tabular}[c]{@{}c@{}}0.1635\\ 0.1643\end{tabular}} &
  \textbf{\begin{tabular}[c]{@{}c@{}}0.0686\\ 0.0779\end{tabular}} &
  \textbf{\begin{tabular}[c]{@{}c@{}}0.0624\\ 0.0643\end{tabular}} &
  \textbf{\begin{tabular}[c]{@{}c@{}}0.0242\\ 0.0407\end{tabular}} &
  \textbf{\begin{tabular}[c]{@{}c@{}}0.0204\\ 0.0266\end{tabular}} \\
  \hline
\%Improv. &
  \begin{tabular}[c]{@{}c@{}}@10\\ @20\end{tabular} &
  \begin{tabular}[c]{@{}c@{}}47.4\%\\ 23.4\%\end{tabular} &
  \begin{tabular}[c]{@{}c@{}}13.7\%\\ 12.1\%\end{tabular} &
  \begin{tabular}[c]{@{}c@{}}28.2\%\\ 8.16\%\end{tabular} &
  \begin{tabular}[c]{@{}c@{}}22.8\%\\ 11.2\%\end{tabular} &
  \begin{tabular}[c]{@{}c@{}}36.4\%\\ 2.23\%\end{tabular} &
  \begin{tabular}[c]{@{}c@{}}44.1\%\\ 17.3\%\end{tabular} &
  \begin{tabular}[c]{@{}c@{}}10.5\%\\ 7.39\%\end{tabular} &
  \begin{tabular}[c]{@{}c@{}}3.03\%\\ 3.50\%\end{tabular} \\ \hline
  \end{tabular}
\end{table*}

\begin{table*}[t]

  \caption{Results of Recall@ and NDCG@ for GNAQ vs. Baseline Models with 256-dimensional Embeddings.}
  \label{tab:tab6dimensional256}
  \begin{tabular}{cccccccccc}
    \hline
\multirow{2}{*}{Model} &
   &
  \multicolumn{2}{c}{MovieLens} &
  \multicolumn{2}{c}{Gowalla} &
  \multicolumn{2}{c}{Yelp2020} &
  \multicolumn{2}{c}{Amazon-Book} \\ \cline{3-10} 
 &
  Metric &
  Recall &
  NDCG &
  Recall &
  NDCG &
  Recall &
  NDCG &
  Recall &
  NDCG \\ \hline

LightGCN~\cite{he2020lightgcn} &
  \begin{tabular}[c]{@{}c@{}}@10\\ @20\end{tabular} &
  \begin{tabular}[c]{@{}c@{}}0.1708\\ 0.2641\end{tabular} &
  \begin{tabular}[c]{@{}c@{}}0.2904\\ 0.2988\end{tabular} &
  \begin{tabular}[c]{@{}c@{}}0.1305\\ 0.1833\end{tabular} &
  \begin{tabular}[c]{@{}c@{}}0.1142\\ 0.1301\end{tabular} &
  \begin{tabular}[c]{@{}c@{}}0.0548\\ 0.0887\end{tabular} &
  \begin{tabular}[c]{@{}c@{}}0.0386\\ 0.0496\end{tabular} &
  \begin{tabular}[c]{@{}c@{}}0.0261\\ 0.0446\end{tabular} &
  \begin{tabular}[c]{@{}c@{}}0.0217\\ 0.0286\end{tabular} \\
  \hline
BiGeaR~\cite{chen2022learning} &
  \begin{tabular}[c]{@{}c@{}}@10\\ @20\end{tabular} &
  \begin{tabular}[c]{@{}c@{}}\underline{0.1631}\\ 0.1823\end{tabular} &
  \begin{tabular}[c]{@{}c@{}}\underline{0.4051}\\ \underline{0.3879}\end{tabular} &
  \begin{tabular}[c]{@{}c@{}}\underline{0.1283}\\ 0.1622\end{tabular} &
  \begin{tabular}[c]{@{}c@{}}\underline{0.1400}\\ \underline{0.1551}\end{tabular} &
  \begin{tabular}[c]{@{}c@{}}\underline{0.0550}\\ \underline{0.0811}\end{tabular} &
  \begin{tabular}[c]{@{}c@{}}\underline{0.0475}\\ \underline{0.0597}\end{tabular} &
  \begin{tabular}[c]{@{}c@{}}\underline{0.0240}\\ \underline{0.0417}\end{tabular} &
  \begin{tabular}[c]{@{}c@{}}0.0199\\ \underline{0.0258}\end{tabular} \\
N2UQ~\cite{liu2022nonuniform} &
  \begin{tabular}[c]{@{}c@{}}@10\\ @20\end{tabular} &
  \begin{tabular}[c]{@{}c@{}}0.1340\\ \underline{0.2091}\end{tabular} &
  \begin{tabular}[c]{@{}c@{}}0.2952\\ 0.2928\end{tabular} &
  \begin{tabular}[c]{@{}c@{}}0.1235\\ \underline{0.1707}\end{tabular} &
  \begin{tabular}[c]{@{}c@{}}0.1344\\ 0.1438\end{tabular} &
  \begin{tabular}[c]{@{}c@{}}0.0375\\ 0.0606\end{tabular} &
  \begin{tabular}[c]{@{}c@{}}0.0427\\ 0.0493\end{tabular} &
  \begin{tabular}[c]{@{}c@{}}0.0225\\ 0.0392\end{tabular} &
  \begin{tabular}[c]{@{}c@{}}\underline{0.0238}\\ 0.0304\end{tabular} \\
GPTQ~\cite{frantaroptq} &
  \begin{tabular}[c]{@{}c@{}}@10\\ @20\end{tabular} &
  \begin{tabular}[c]{@{}c@{}}0.1097\\ 0.1733\end{tabular} &
  \begin{tabular}[c]{@{}c@{}}0.0970\\ 0.1184\end{tabular} &
  \begin{tabular}[c]{@{}c@{}}0.0624\\ 0.0918\end{tabular} &
  \begin{tabular}[c]{@{}c@{}}0.0649\\ 0.0740\end{tabular} &
  \begin{tabular}[c]{@{}c@{}}0.0208\\ 0.0350\end{tabular} &
  \begin{tabular}[c]{@{}c@{}}0.0175\\ 0.0226\end{tabular} &
  \begin{tabular}[c]{@{}c@{}}0.0132\\ 0.0222\end{tabular} &
  \begin{tabular}[c]{@{}c@{}}0.0141\\ 0.0175\end{tabular} \\
LEGCF~\cite{liang2024lightweight} &
  \begin{tabular}[c]{@{}c@{}}@10\\ @20\end{tabular} &
  \begin{tabular}[c]{@{}c@{}}0.0784\\ 0.1277\end{tabular} &
  \begin{tabular}[c]{@{}c@{}}0.1593\\ 0.1586\end{tabular} &
  \begin{tabular}[c]{@{}c@{}}0.0839\\ 0.1222\end{tabular} &
  \begin{tabular}[c]{@{}c@{}}0.0704\\ 0.0822\end{tabular} &
  \begin{tabular}[c]{@{}c@{}}0.0268\\ 0.0467\end{tabular} &
  \begin{tabular}[c]{@{}c@{}}0.0191\\ 0.0255\end{tabular} &
  \begin{tabular}[c]{@{}c@{}}0.0162\\ 0.0283\end{tabular} &
  \begin{tabular}[c]{@{}c@{}}0.0139\\ 0.0183\end{tabular} \\
\textbf{GNAQ (Ours)} &
  \begin{tabular}[c]{@{}c@{}}@10\\ @20\end{tabular} &
  \textbf{\begin{tabular}[c]{@{}c@{}}0.2353\\ 0.2362\end{tabular}} &
  \textbf{\begin{tabular}[c]{@{}c@{}}0.5370\\ 0.4288\end{tabular}} &
  \textbf{\begin{tabular}[c]{@{}c@{}}0.1698\\ 0.1815\end{tabular}} &
  \textbf{\begin{tabular}[c]{@{}c@{}}0.1793\\ 0.1781\end{tabular}} &
  \textbf{\begin{tabular}[c]{@{}c@{}}0.0780\\ 0.0869\end{tabular}} &
  \textbf{\begin{tabular}[c]{@{}c@{}}0.0703\\ 0.0720\end{tabular}} &
  \textbf{\begin{tabular}[c]{@{}c@{}}0.0297\\ 0.0496\end{tabular}} &
  \textbf{\begin{tabular}[c]{@{}c@{}}0.0243\\ 0.0318\end{tabular}} \\
  \hline
\%Improv. &
  \begin{tabular}[c]{@{}c@{}}@10\\ @20\end{tabular} &
  \begin{tabular}[c]{@{}c@{}}44.3\%\\ 12.9\%\end{tabular} &
  \begin{tabular}[c]{@{}c@{}}32.6\%\\ 10.5\%\end{tabular} &
  \begin{tabular}[c]{@{}c@{}}32.3\%\\ 6.33\%\end{tabular} &
  \begin{tabular}[c]{@{}c@{}}28.1\%\\ 14.8\%\end{tabular} &
  \begin{tabular}[c]{@{}c@{}}41.8\%\\ 7.15\%\end{tabular} &
  \begin{tabular}[c]{@{}c@{}}48.0\%\\ 20.6\%\end{tabular} &
  \begin{tabular}[c]{@{}c@{}}23.8\%\\ 18.9\%\end{tabular} &
  \begin{tabular}[c]{@{}c@{}}2.10\%\\ 23.3\%\end{tabular} \\ \hline
  \end{tabular}
\end{table*}

\begin{table*}[t]
  \caption{Parameter sizes of different quantization methods and training time (in seconds) under the 256-dimensional encoding.}
  \label{tab:tab7ParameterSize}
  \begin{tabular}{cccccc}
    \toprule
     Model  & Number   of parameters  & MovieLens & Gowalla & Yelp2020 & Amazon-Book \\
     \midrule
    BiGeaR~\cite{chen2022learning} & $O\left(N\left(L\:+\:1\right)\left(32\:+\:d\right)\right)$ & 14494     & 33950   & 61217    & 298037      \\
N2UQ~\cite{liu2022nonuniform}   & $O\left(2N\left(32\:+\:d\right)\right)$        & 2188      & 16002   & 29162    & 32604  \\
GPTQ~\cite{frantaroptq}   & $O\left(4N\left(32\:+\:d\right)\right)$        & 32        & 119     & 332      & 1069   \\
LEGCF~\cite{liang2024lightweight}  & $O(500d+2N)$         & 6780      & 34823   & 37904    & 75916       \\
\textbf{GNAQ (Ours)}& $O\left(N\left(d\:+\:128\right)\right)$        & 6880      & 14815   & 41765    & 133482  \\ 
     \bottomrule
  \end{tabular}
\end{table*}

\begin{table}[t]
\small
  \caption{Ablation Experiments 
  }
  \label{tab:tab8Ablation}
  \begin{tabular}{ccccc}
    \toprule
     \multicolumn{1}{c}{\multirow{2}{*}{Model}} & \multicolumn{2}{c}{Gowalla} & \multicolumn{2}{c}{Yelp2020} \\ \cline{2-5} 
                       & Recall@20     & NDCG@20     & Recall@20      & NDCG@20     \\
     \midrule
    GNAQ                   & 0.1750        & 0.1643      & 0.0779         & 0.0643      \\ \hline
w/o DQS                & 0.1613        & 0.1263      & 0.0708         & 0.0431      \\
w/o RAU                & 0.1424        & 0.0973      & 0.0518         & 0.0274      \\
w/o rank\_loss         & 0.1689        & 0.1245      & 0.0740         & 0.0432      \\ 
     \bottomrule
  \end{tabular}
\end{table}

\subsection{Overall Performance Analysis (RQ1)}

Extensive experiments were conducted on four datasets with different embedding dimensions to compare our approach against the baseline models. The results are presented in Tables~\ref{tab:tab4dimensional64} to~\ref{tab:tab6dimensional256}. 

    \textbf{Superior Performance of GNAQ:} Across almost all evaluation metrics on four datasets, our approach outperforms other baseline models. For example, on the Gowalla dataset (Figure~\ref{fig:fig4Gowalla}), GNAQ achieves the highest Recall@10, Recall@20, and NDCG@10 across all embedding dimensions, except for a slight NDCG@20 deficit compared to BiGeaR at 64 dimensions. It may be because of insufficient information from neighbors, the Role of Relational Aggregation-Based Update (RAU) is limited in its effect during the quantization coding update. The average improvements in Recall@10, Recall@20, NDCG@10, and NDCG@20 are 27.8\%, 10.2\%, 17.6\%, and 8.51\%, respectively, demonstrating its effectiveness in preserving recommendation accuracy.
  
  \textbf{Impact of Embedding Dimensions:} As the embedding dimension increases, our approach demonstrates stronger performance relative to baseline models. Higher-dimensional embeddings contain more information, and GNAQ’s dynamic quantization step strategy better captures inter-dimensional correlations in node embeddings. For instance, on the Yelp2020 dataset, GNAQ’s NDCG@10 at 256 dimensions is 48\% higher than GPTQ, highlighting its advantage in high-dimensional scenarios.
  
  \textbf{Sparsity-Dependent Performance:} Our approach achieves more significant improvements on datasets with lower sparsity (e.g., MovieLens). Dense interaction graphs in such datasets contain richer high-order interaction information, which GNAQ leverages through dynamic quantization steps to adapt to node embedding distributions, thereby enhancing recommendation quality.

\subsection{Resource Consumption Analysis (RQ2)}

Table~\ref{tab:tab7ParameterSize} compares the parameter counts and training times of different methods. Key observations include:

\textbf{Efficiency of Direct Optimization Methods:} Directly optimizing quantized  based models (e.g., GNAQ, N2UQ) exhibit lower training times compared to knowledge distillation-based approaches (e.g., BiGeaR). For example, GNAQ’s training time on the Amazon-Book dataset is 133,482 seconds, significantly less than BiGeaR’s 298,037 seconds, as it avoids re-training full-precision models.

\textbf{Balance Between Efficiency and Effectiveness:} GNAQ’s parameter size is higher than that of lightweight models (e.g. LEGCF), but achieves superior performance in terms of Recall and NDCG. Compared to BiGeaR, GNAQ uses fewer parameters while outperforming it in most scenarios, demonstrating a better trade-off between efficiency and effectiveness.

\subsection{Ablation Experiments (RQ3)}
Ablation studies are conducted on two datasets at 128 embedding dimensions to analyze the contributions of 
dynamic quantization steps (DQS), relational aggregation-based update (RAU), and ranking-enhanced loss (rank-loss), shown in Table~\ref{tab:tab8Ablation}.

\textbf{Effectiveness of Dynamic Quantization Steps (DQS):} Removing DQS (w/o DQS) leads to a significant performance degradation (e.g., Gowalla's NDCG @ 20 drops from 0.1643 to 0.1263), confirming that node-aware dynamic quantization steps effectively adapt to embedding distributions and reduce information loss.

\textbf{Role of Relational Aggregation-Based Update (RAU):} Removing RAU (w/o RAU) causes more severe performance declines (e.g., Yelp2020’s Recall@20 drops to 0.0518), indicating that graph interaction-aware embedding updates are critical for quantization accuracy by guiding parameter updates with structural information.

\textbf{Impact of Ranking-Enhanced Loss (rank-loss):} Removing rank-loss (w/o rank-loss) reduces NDCG@20 while preserving Recall@20, demonstrating its specific contribution to improving ranking accuracy without compromising recall, thus improving overall quality of recommendation.

\section{Conclusion and Future Work}
This paper proposes a novel graph node-aware quantization with dynamic step size for collaborative filtering. 
Our GNAQ effectively captures the structural heterogeneity of graph data and achieves superior performance over state-of-the-art quantization methods. Extensive experiments on four public datasets demonstrate the effectiveness and efficiency of our approach. In the future, 
we apply GNAQ to actual edge deployment scenarios to achieve efficient inference on resource-constrained devices while maintaining high recommendation performance.

\section*{Acknowledgments}
This work is partially supported by NSFC, China (No.62276196) and JSPS Invitational Fellowships for Research in Japan (Short-term).

\section*{Usage of Generative AI}
The general outline of the entire article was independently completed by us and Generative AI was used to modify the grammar of the sentences and polish the expression of the language.

\bibliographystyle{ACM-Reference-Format}
\bibliography{references}

\end{CJK}
\end{document}